# Unlensing Multiple Arcs in 0024+1654: Reconstruction of the Source Image

Wesley N. Colley [1] [2], J. Anthony Tyson [3], Edwin L. Turner [4]

## ABSTRACT

A unique reconstruction of the image of a high redshift source galaxy responsible for multiple long arcs in the $z = 0.4$ cluster 0024+1654 is obtained by inverse lensing. Deep B and I imaging with the *Hubble Space Telescope*[5] enables high resolution of the arcs due to strong gravitational lensing of the background source. The gravitational lens in the foreground cluster is thus used to obtain a magnified view of the distant source. Four strongly lensed images of the source lead to a unique reconstruction. Each of the long arcs, when unlensed, leads to the same reconstructed source image exhibiting a beaded ring-like morphology. The U luminosity of the ring alone is equivalent to a normal galaxy. This is likely a galaxy in formation.

*Subject headings:* gravitational lensing — cosmology: observations — methods: observational — clusters — observation, specific Cl0024+1654 — galaxies: formation

## 1. Introduction

Strong gravitational lensing can in principle provide unprecedented views of distant background sources (Zwicky 1937; see Blandford & Narayan 1992 for a review). This can lead to a unique reconstruction of the source image, given enough information in the lensed image; generally this works only if there are multiple images of the source. Kochanek & Narayan (1992) obtained a full inversion of a lensed radio source in both intensity and polarization, the full Einstein ring providing the needed constraints. Even when there is insufficient data in the image plane to construct a unique inversion, useful information on resolved features in the source may be obtained. A recent example is the resolution of the z = 2.29 infrared source FSC10214+4724 via gravitational lensing due to a foreground galaxy (Eisenhardt *et al.* 1996). The increase in source resolution provided by such single galaxy lenses is limited by the $\approx 1''$ deflection, due to the small magnification.

---

[1] Supported by the Fannie and John Hertz Foundation, Livermore, CA 94551-5032

[2] Princeton University Department of Astrophysical Sciences, Princeton, NJ 08544 Email: wes@astro.princeton.edu

[3] AT&T Bell Labs, 600 Mountain Ave., Murray Hill, NJ 07974 Email:tyson@physics.att.com

[4] Princeton University Department of Astrophysical Sciences, Princeton, NJ 08544 Email: elt@astro.princeton.edu

[5] Based on observations with the NASA/ESA *Hubble Space Telescope*, obtained at the Space Telescope Science Institute, which is operated by AURA, under NASA contract NAS 5-26555

In principle, galaxy cluster lenses can yield magnifications 30 times larger. Since the discovery of strongly lensed long "arcs" around massive clusters of galaxies (Soucail *et al.* 1987; Lynds & Petrosian 1986) it has been hoped that such highly stretched images of background galaxies could be used to get at least a one-dimensional improvement of resolution of the source galaxy. Indeed, several long arcs seen in ground-based imaging show structure along the arc. The $0.1''$ resolution of the HST, combined with the magnification by the gravitational lens, can yield resolution of $\approx 0.01''$ in the source plane.

The reconstructed source image is however only as accurate as the lens model. Thus, one concern has been whether such a reconstruction is unique. This is particularly worrysome in cases where there are few constraints on the detailed lens model, as in single long arcs. The multiple long arcs in the z = 0.39 cluster 0024+1654 present an opportunity to break some lens model degeneracies and develop a unique reconstruction of the source. These relatively bright multiple arcs in 0024+1654 were discovered by Koo (1988) photographically and various lens models based on deeper ground-based CCD imaging of these arcs have been advanced (Kassiola *et al.* 1992; Wallington *et al.* 1995). Deep imaging suggests at least three and perhaps four strongly lensed images of the same blue source. These models however were based either on assuming only three source images and/or an incomplete search of the source plane.

Five arcs, images of the same blue source galaxy, are clearly seen in the HST imaging (below). The same detailed morphology is seen in each arc, but flipped by the lens. Thus it is possible to sufficiently constrain a full lens model in which the parameters for the dark matter lens and cluster galaxy lenses are all allowed to vary. In this *Letter* we present a fully two-dimensional reconstruction with 100 pc resolution for the high redshift source of the major arcs seen in the distant massive cluster 0024+1654.

## 2. Observations and Data Reduction

The data were taken on 14 October 1994 UT in three visits of two orbits each, plus one additional orbit. The total integration time was 15000 sec. We obtained 3 series of orbits in each of two filters: F450W (blue) and F814W (red). The 1100 sec exposures were offset by approximately $10''$ enabling us to fill in the gaps between the WFCP2 chips effectively, and cover slightly more sky at the cost of some integration time at the edges of the exposed sky area.

We made multiple passes on the data, using an initial cleaned coadd of all the data as support for bad pixel removal. After cosmic ray detection and removal using STSDAS *crrej*, we coadded the data for each of the three orbital series. Each of the three stacks represents seven images, totalling 8400 seconds in blue, and six images totalling 6600 seconds in red. We then registered the images for each chip from the three orbital series and obtained their mean, excluding areas with no data. This process yielded a stacked image of the three runs, most of which contained the full signal from all three runs, with the edges having 30% less exposure. Next, we performed a final cleaning of the six images (three chips, two bands), first using *daofind* and *imclean* to find and remove bad or hot pixels, then manually removing them with *imedit*.

We then mosaicked the three chips (again employing IRAF *geotran* for registration) to obtain a continuous image in the familiar "L" configuration of WFC2 measuring $2.7'$ on a side. We have not included the data recorded by the Planetary Camera on the same run, due to its bright surface brightness limit. To produce a pseudo-color image, we sky-subtracted and calibrated the blue and red images. We then produced a third intermediate band by mean-averaging the blue and red band images. These three RGB images, converted to stretched 8 bit log intensity, were then packed into a $3 \times 8$ bit color tiff image shown in three quadrants of Figure 1 [Plate 000].

Figure 2 is a stretched monochrome reproduction of part of the blue image and shows our naming convention for the five arcs. Note in the color WFPC2 image shown in Figure 1 that arcs A, B, and C as well as a counter-arc D and a demagnified subimage E can be identified by their identical morphology, blue color, and surface brightness (preserved by lensing). Lensed images at these positions were also found by experiments with forward ray tracing a single diffuse source through a complete lens model to be described elsewhere.

## 3. Source Reconstruction

The WFPC2 image in Figure 1 shows that the arcs are well resolved in these images, revealing at least 5 separate features in the common source. Arcs A,C, and D are highly magnified, and we concentrate on producing a self-consistent source image reconstruction from these three arcs. We also use the location and orientation of arcs B and E to constrain our multiple lens parameter space search. Our procedure for image reconstruction is somewhat different than usually followed in lens modeling. Rather than seeking a unique lens model, we instead make use of the multiple images to arrive at a self-consistent source image reconstruction by varying the parameters of the multiple lenses. Some lens parameters are separately constrained by the known location of cluster galaxies and from arclet data over wider field deep ground-based imaging.

### 3.1. Lens model

Our goal is to place the reconstructions of arcs A,C and D at nearly the same point in the source plane, produce the correct image parity, and generate nearly identical orientation and dimension in all three reconstructions. We begin with a softened isothermal central dark matter distribution, which produces a deflection angle (Grossman & Narayan 1988)

$$\alpha = \left\{ \begin{array}{ll} \alpha_0 \cdot (r/r_{core}) \left(0.75 - r^2/8r_{core}^2\right) & r < r_{core} \\ \alpha_0 \left(1 - 3r_{core}/8r\right) & r \geq r_{core} \end{array} \right\} \quad (1)$$

where $\alpha_0 = 4\pi\sigma_v^2/c^2$, the deflection angle for an isothermal distribution. All lenses in the cluster are modelled as soft core isothermal distributions with an outer mass cutoff, giving (for the spherical case) a deflection angle

$$\alpha = \left\{ \begin{array}{ll} \alpha_0 \cdot (r/r_{core}) \left(0.75 - r^2/8r_{core}^2\right) & r < r_{core} \\ \alpha_0 \left\{(5r_{core}/8r) + (\beta r_{out}/r) \left[\arctan(r/r_{out}) - \arctan(r_{core}/r_{out})\right]\right\} & r \geq r_{core} \end{array} \right\} \quad (2)$$

where $r_{core}$ is the soft core radius, $r_{out}$ is the outer mass cutoff, and $\beta = (1 + r_{core}^2/r_{out}^2)$.

The lenses consist of the massive dark matter lens and the many cluster galaxies. The parameters for each lens are its location on the sky, its effective velocity dispersion, ellipticity, orientation, and outer cutoff. Dressler & Gunn (1992) have studied this cluster and list photometry and redshifts for many of the brighter cluster galaxies. We computed approximate Faber-Jackson (FJ) masses for 48 of the brighter lens galaxies based on aperture photometry.

The positions of the cluster galaxies are fixed by their observed red centroids, and the orientation of their mass ellipticity is set at the observed isophote orientation. Rough starting values for their velocity dispersions are obtained from the FJ relation, after K-correction, using the observed 814nm (760 nm in cluster rest frame) magnitudes and B–I and R–I colors: $Log\sigma_v = 4.3 - R/10$ ($h = 0.7$).

### 3.2. Unlens

The starting parameters for the main dark matter lens were obtained from weak lens inversion of 1200 arclets, *not* the strong lensed arcs, from deep wide-field 4-meter imaging (described elsewhere). Rather than forward ray-trace a model source image, we developed code to backward ray-trace the observed arcs to the source. First, a mapping from the image plane to the source plane is created by vector adding the pretabulated radial distortions of the lens components. To ensure adequate resolution in the source plane, the program dynamically subsamples the image plane in small triangles. Each triangular sub-pixel maps into a triangle on the source plane. Source plane pixels inside (or on the border of) this triangle are filled according to the fraction of that pixel occupied by the triangle. Count is kept of the total amount of image plane mapped into each source plane pixel, necesary for normallization at the end of the sequence. To speed calculation, there is a lower threshold for image pixels.

We found results from this scheme to agree with an equivalent and independent code we haved also developed, called *unlens*. This code uses multiple ray pixel subsampling of the image plane, rather than polygon filling to generate the source image, but is otherwise similar to the first algorithmically. However, the second code is much more flexible in handling different input parameters, lens models and image plane data. As such, it will be presented for general use in a future paper (Tyson 1996).

### 3.3. Optimization

Since these lens masses are approximate and do not account for some of the halo dark matter in each cluster galaxy, we vary the outer cutoff of some of the galaxy masses, particularly those projected nearby any of the arcs. Our goodness of fit criterion is based on an optimal match for the reconstructions of arcs A, C and D in the source plane. Since we have 49 lenses in our model, with six parameters each, we could have spent a great deal of time searching through the available 294-dimensional space to optimize our results. Instead, we initially tried to optimize our reconstructions as a function of the location, velocity dispersion, and ellipticity and orientation of the central dark potential, keeping the cluster galaxy mass parameters fixed at their FJ values. We found that the reconstructed position in the source plane of different arcs depends strongly and *differently* on the lens parameters of the main potential as well as several individual cluster galaxy parameters.

Figures 3 (a,b,c,d) illustrate our method. We selected the three prominent features seen in all three arcs (see Plate 1), and followed these features through the lens back to the source plane for each arc. In each of these figures we plot two graphs: the lower shows the trajectory of the reconstructed source features in the source plane (1 pixel = $0.025''$), while the upper plot gives the rms error in reconstructed source image positions in the source plane as a function of the parameter varied: main potential x or y position in the image plane [1 pixel = $0.1''$], main potential velocity dispersion, main potential ellipticity. Figure 3a shows the trajectories of these features in the reconstructed source as they respond to movement of the main potential centroid in along $x$. Similarly, Figures 3(b,c,d) show those trajectories in response to b) movement of the main potential in $y$, c) variation of its velocity dispersion, and d) variation of the ellipticity of the main dark matter potential.

The "optimal" fit is given as the three roughly overlapping triangles near the center. Radiating from each triangle are the tracks followed by each of the triangles responding to the variable in question. The tracks extend to arbitrary end-points, which are denoted again by heavy triangles. A global mix of these rms differences is what we want to minimize in order to reach self-consistency in the source reconstruction. Finally, we found that six cluster galaxies which are projected near the arcs influence the shape of the reconstructed source for that arc. We varied the lens parameters for these six galaxies from the starting FJ values to improve the match between the reconstructed images for arcs A, C, and D using the squared difference of their images as an error criterion. This procedure was repeated recursively, taking various routes through the lenses, to arrive at a global optimization.

### 3.4. Source image

Our reconstructions of the source image corresponding to arcs A, C, and D are shown in the lower left corner of Plate 1. These pseudo-color images were generated as described in section 2. In each image, the image plane WFPC2 pixelization is visible and distorted in a different way for each of the three images, as expected. However, the main features of the source appear to be produced self-consistently in each plot. Both the noise and artifacts are damped when the images are coadded, as shown in the last of the four reconstructed source images in Plate 1. The similarity of the coadded image and the reconstructions to each other gives us some confidence in the reconstruction.

Rather than summing the three reconstructed arcs, we may demand complete morphological consistency by instead combining them nonlinearly, so that pixels must be consistent in two out of three of the reconstructed images. The harmonic mean of the three reconstructed source images is shown in monochrome in Figure 4a. In Figures 4b and 4c, we show for comparison rotated and magnified images of arcs B and E. Note the morphological similarity between unreconstructed arcs B and E, and the reconstructed source image from arcs A, C, and D. We do not unlens arcs B and E here, due to the extreme sensitivity to the mass distribution of the nearby galaxies. This will be discussed elsewhere.

## 4. Analysis of the Source

The reconstructed source resembles a blue "theta" or "B" of size $1''$. With its beaded ring-like nature, the source galaxy does not lie along the Hubble sequence. Moreover, it is large by comparison to most faint blue galaxies found in the HST Medium Deep Survey (Im et al. 1995). Most galaxies this bright with redshifts over $z = 1$ have angular diameters about two times smaller (Mutz et al. 1994). However, if the source were just slightly farther away, the $(1 + z)^4$ dimming would render all but the bright knots invisible and thus artificially reduce the apparent size. The source is likely to lie at a redshift between 1.2 and 1.8, since no emission line has been seen in deep spectroscopy. With a corresponding angular diameter distance of 860–890 $h^{-1}$ Mpc, $1''$ subtends 4.2–4.3 $h^{-1}$ kpc diameter.

Over this source redshift range, the F450W filter corresponds to 160–205nm in the source rest-frame, and the F814W filter moves to 290–370nm (U band) in the source rest-frame. This systematic change in morphology from "normal" to "disturbed" seen in HST deep imaging surveys is correlated with redshift, with most of the disturbed galaxies at redshift greater than 1 (Forbes et al. 1995; Schade et al. 1995). We are seeing these distant galaxies at UV wavelengths, highlighting lumpy regions of high star formation. Its very blue color transformed to the rest frame is similar to that found for rapid star forming galaxies (Bruzual & Charlot 1993).

It is interesting to speculate on the origin of the dark core inside the ring and the feature near the center. A very dusty protogalaxy could have this appearance; one would then be seeing UV from HII regions associated with star formation on the outskirts of the dust. Recently a nearby example of a beaded nuclear ring in a barred galaxy has been found in a galaxy undergoing rapid star formation (Buta et al. 1995); a 5 kpc diameter oval ring with bright knots is seen in $H_\alpha$. Other examples of clumpy nuclear rings are known (Morgan 1958; Schommer et al. 1988; Wilson et al. 1991) and several have a nuclear bar. Such a nuclear bar may be visible in Figure 4.

Combining this photometry with deep ground-based BRIgri photometry and calibrating on the gri system, the knots around the ring in the source galaxy have an effective g-i color of -0.4 to 0.2 – characteristic of the bluest galaxies known. The B-R and r-i colors are similar to the colors of the faint blue galaxy population. The central "bar" in the source has a B-I color 0.7 mag redder than the mean color of 3 blue knots around the ring. This lends support to the dust hypothesis; taking a nearby nuclear ring galaxy as an example, (see Buta *et al.* Figure 8) over 3 mag of localized extinction at 200nm would be required to absorb the light interior to 4 kpc. The peak in interstellar dust absorption at 220 nm leads to sufficient extinction to make the observed hole, given a UV color excess of 0.7 mag if we are to assume a rather flat extinction curve (low $R_\lambda$), which is typical of dense molecular regions where the grain sizes are large. If primeval, this object may not yet have formed a normal disk or spheroid, and may consist mainly of a high SFR ring. Any spheroid should become visible in the IR. Unlensing deep JHK-band imaging on the Keck may place more stringent limits on any older population of stars; the H-K break for the source is expected longwards of 860 nm, predicting a relatively red I-J color.

The reconstructed source has an unlensed apparent magnitude of 23.7 i mag, which is consistent with $z > 1.2$ (Schade *et al.* 1995). For $z > 1.2$, its absolute U magnitude is brighter than -20 mag (h = 0.7) from the ring alone. Accounting for several magnitudes of nuclear extinction at 200 nm, would lead to highly luminous source in the IR. The compound microscope formed by the cluster lens and HST affords a rare view of a young star-forming high redshift galaxy.

We thank Phil Fischer, Frank Valdes and Rick Wenk for help debugging *unlens*, and Greg Kochanski for discussions. We also thank Bruce Draine for his useful comments on dust absorption. WNC is most grateful for the continued support of the Fannie and John Hertz Foundation, and would like to thank the Department of Astronomy at the University of Virginia for its hospitality during some of the latter stages of this work. This work was partially funded by NASA grant NAGW-2173.

Fig. 1.— [Plate 1] A pseudo-color image of the field of the massive $z = 0.39$ cluster 0024+1654 generated from coadded 15,000 sec imaging data in the F450W and F814W filters on the HST WFC2 chips. North is up and East is left. The longest side of the image is 2.7 arcminutes. Five blue arcs, each an image of the same background source at much higher redshift may be seen. Three of them, two arcs to the SE and one counter arc to the NW are highly magnified. Our reconstruction of the source [see text] by unlensing each of these longest arcs is shown in the lower left. The source-plane scale is indicated by a 1″ bar.

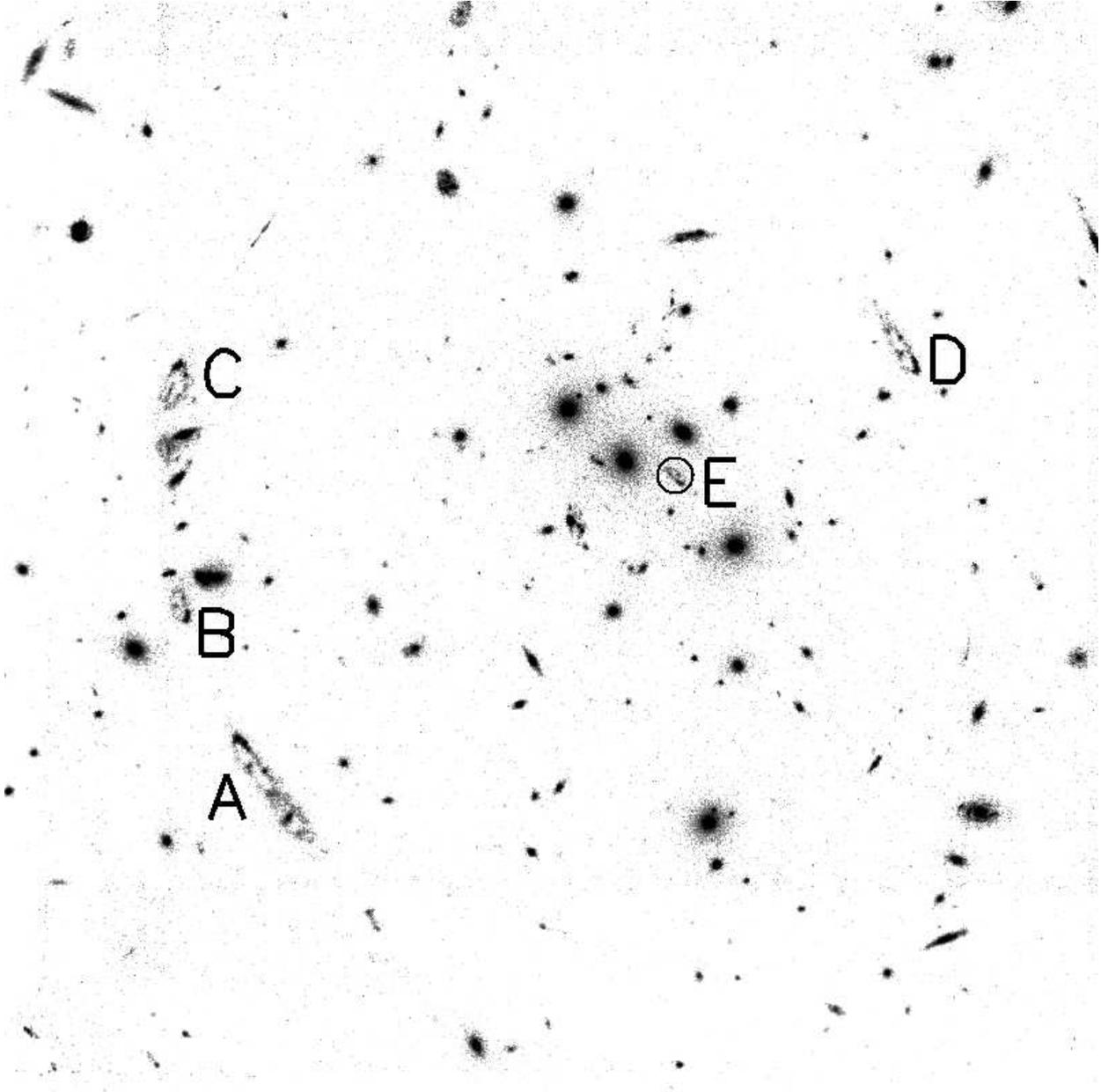

Fig. 2.— Nomenclature for the arcs A, B, C, D, and E marked on a stretched monochrome subfield extracted from the blue WFC2 image. The existence of image E constrains the soft core of the main potential.

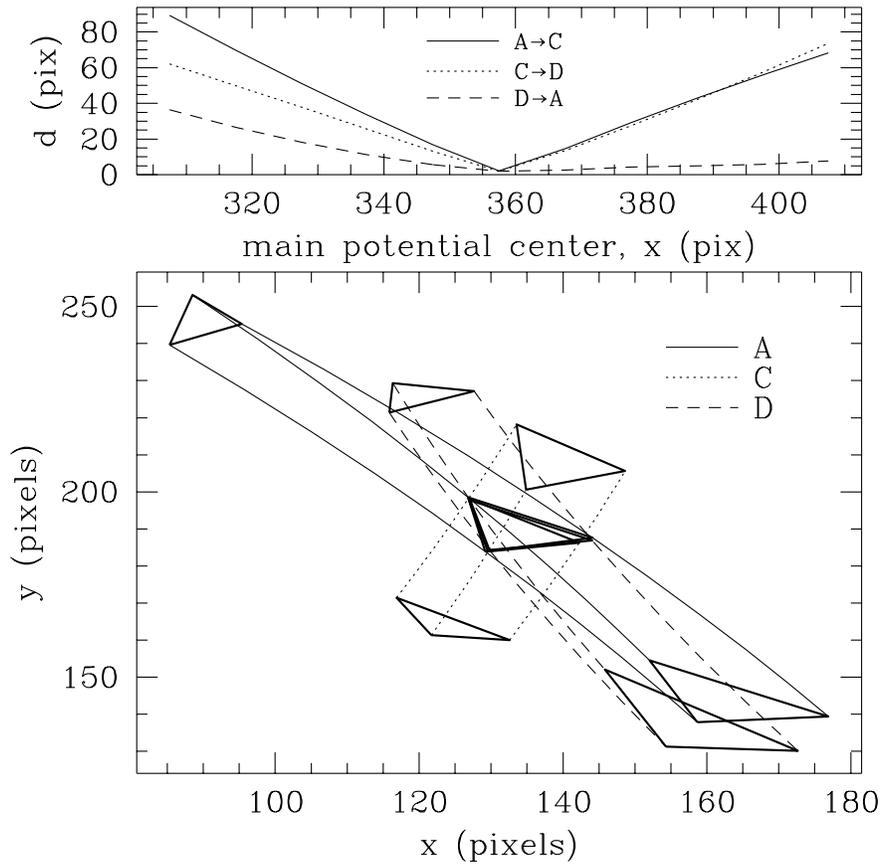

Fig. 3.— a) *bottom*: Trajectories in the source plane of bright features in the reconstructed source image, parameterized in the $x$-position of the main potential, for arcs A,C and D (see text). One pixel is $0.025''$. *top*: RMS distance between the "central" features in the source plane for each of the three reconstructed images.

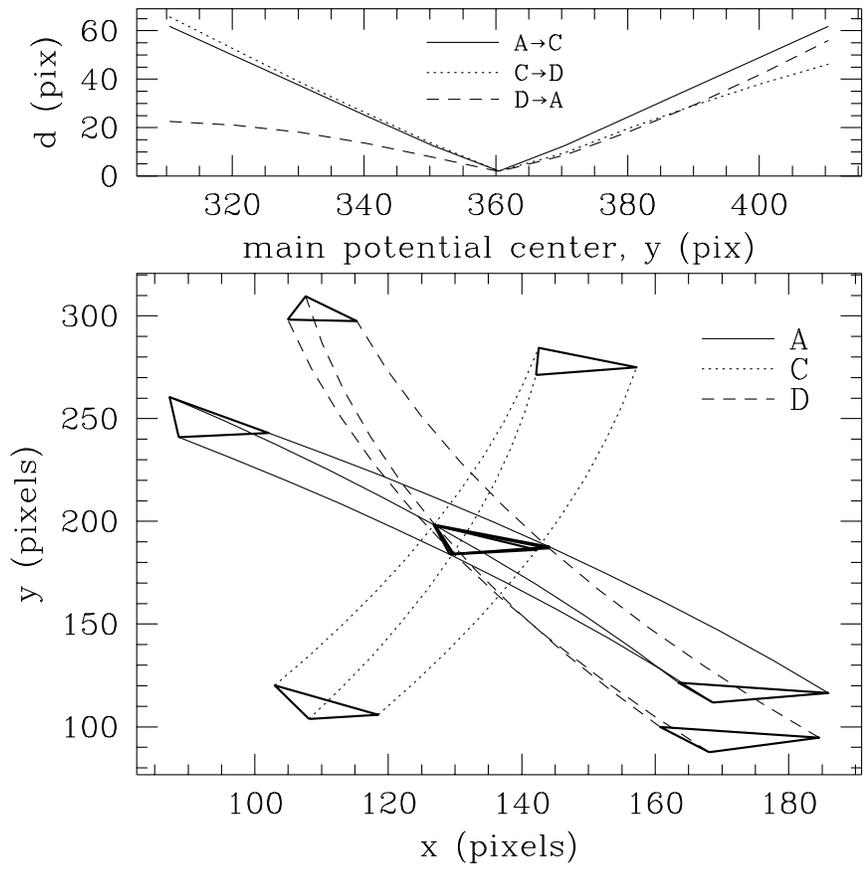

Fig. 3.— b) as in 3a), except the varied parameter is the $y$-position of the main potential.

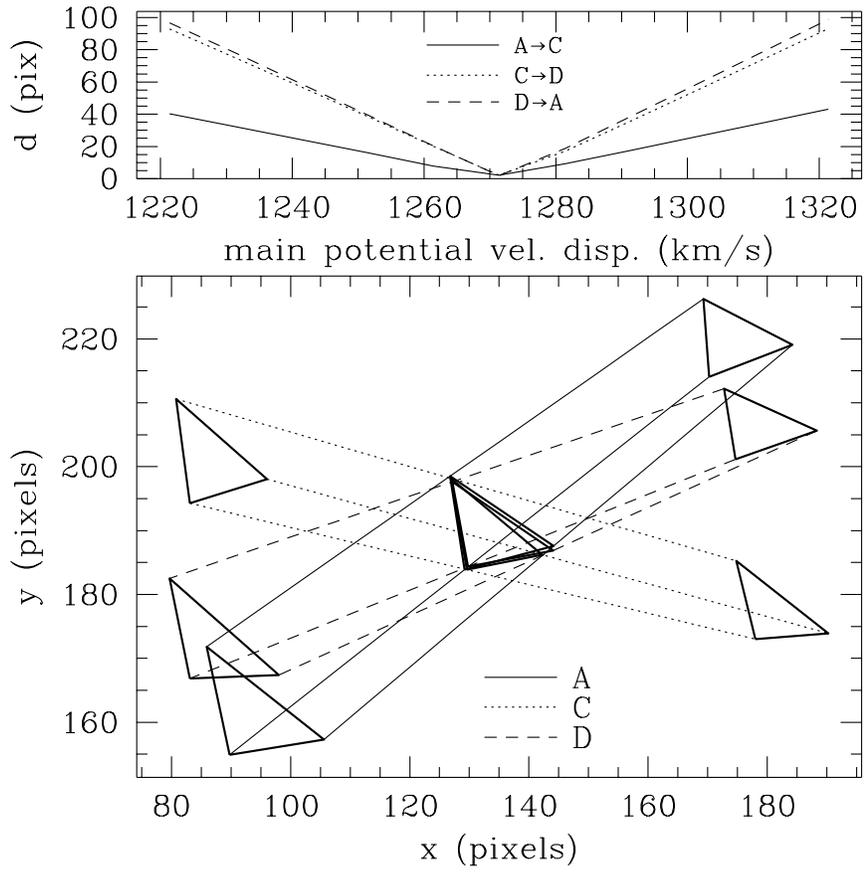

Fig. 3.— c) as in 3a), except the varied parameter is the velocity dispersion of the main potential.

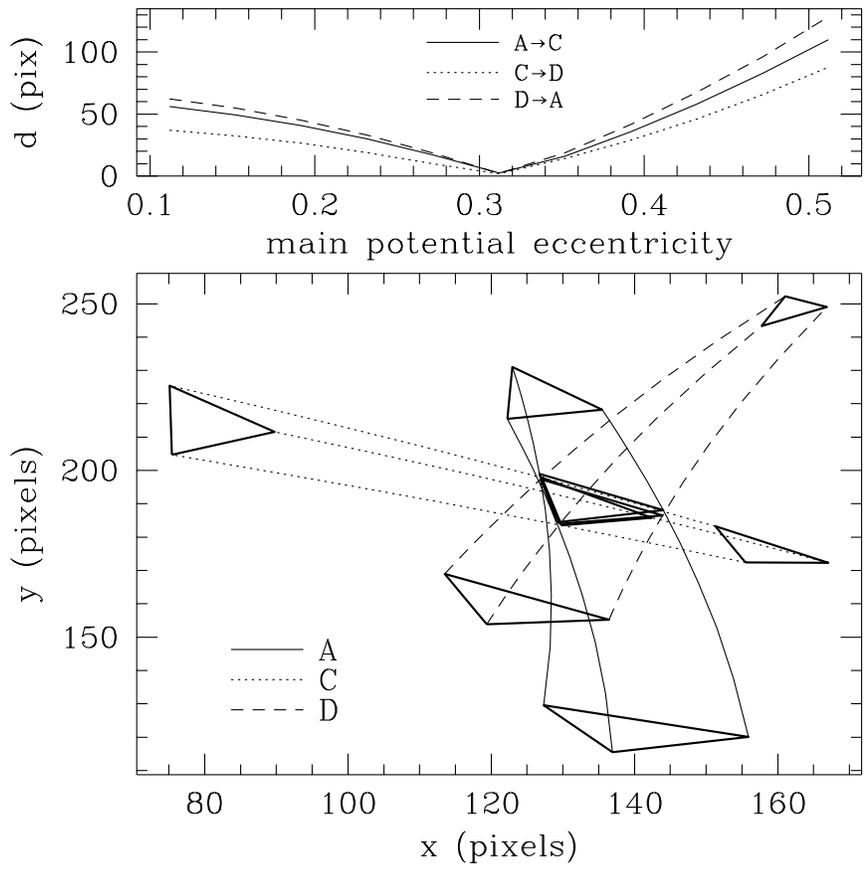

Fig. 3.— d) as in 3a), except the varied parameter is the eccentricity of the main potential.

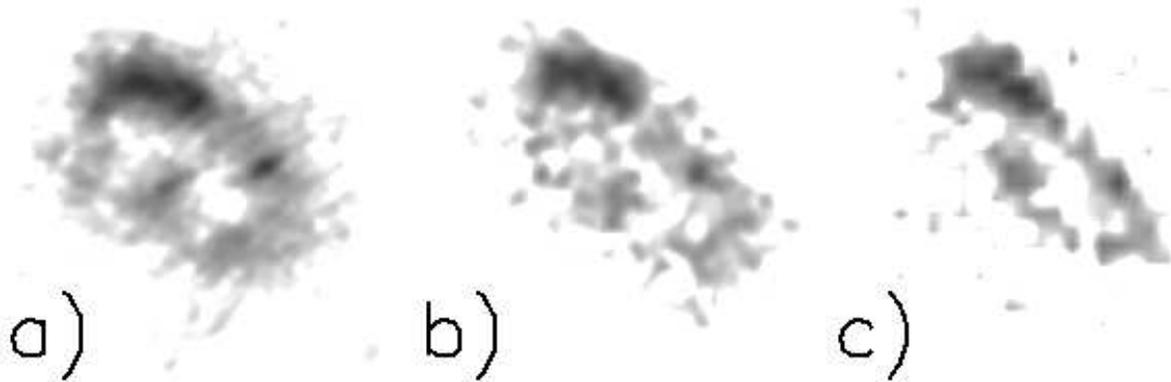

Fig. 4.— a) A reconstructed image of the source galaxy, based on maximum consistency (low pixel reject). b) The un-reconstructed (image plane) image of arc B, rotated and magnified for comparison. c) The un-reconstructed image of arc E, rotated and magnified for comparison.